\documentclass{article}[12pt] 
\usepackage{amssymb} 
\usepackage{amsmath}
\usepackage{amsfonts}
\usepackage{multirow}
\def\be{\begin{equation}}
\def\ee{\end{equation}}
\def\bea{\begin{eqnarray}}
\def\eea{\end{eqnarray}}
\def\aeq#1{\begin{align}#1\end{align}}  
\def\ateq#1#2{\begin{alignat}{#1}#2\end{alignat}}  
\def\Integers{\mathbb{Z}}
\def\thefootnote{\fnsymbol{footnote}}
\begin{document}
\begin{titlepage}
\date{today}       \hfill
{\raggedleft IPMU13-0086}
\begin{center}
\vskip .5in
\renewcommand{\thefootnote}{\fnsymbol{footnote}}
{\Large \bf  Precise lower bound on Monster brane boundary entropy}\\
 
\vskip .250in

\vskip .5in
{\large Daniel Friedan${}^{1,2}$, Anatoly Konechny${^{3,4}}$, and Cornelius Schmidt-Colinet${}^{5}$ }

\vskip 0.5cm
{\it ${}^1$New High Energy Theory Center and Department of Physics and Astronomy,\\
Rutgers, The State University of New Jersey,\\
Piscataway, New Jersey 08854-8019 U.S.A.\\[10pt]

${}^2$The Science Institute, The University of Iceland, Reykjavik, Iceland\\[10pt]
${}^{3}$Department of Mathematics, Heriot-Watt University,\\
EH14 4AS Edinburgh, United Kingdom\\[10pt]

${}^{4}$Maxwell Institute for Mathematical Sciences, Edinburgh, United Kingdom\\[10pt]

${}^{5}$Kavli Institute for the Physics and Mathematics of the Universe,\\
Todai Institutes for Advanced Study (TODIAS),\\
The University of Tokyo, Kashiwa, Chiba 277-8582, Japan}

\end{center}

\vskip .5in

\begin{abstract} \large
In this paper we develop further the linear functional method of deriving lower bounds on the 
boundary entropy of conformal boundary conditions in 1+1 dimensional conformal field 
theories (CFTs). We show here how to use detailed knowledge of the bulk CFT spectrum. 
Applying the method to the Monster CFT  with $c=\bar c=24$ we 
  derive a lower bound $s > - 3.02 {\times} 10^{-19} $
on the boundary entropy $s=\ln g$,
and find
compelling evidence that the optimal bound is $s\ge 0$.
We show that all $g{=}1$ branes
must have the same low-lying boundary spectrum,
which matches the spectrum of the known $g{=}1$ branes,
suggesting that the known examples comprise all possible $g{=}1$ branes,
and also suggesting that the bound $s\ge 0$ holds not just for critical boundary conditions
but for all boundary conditions in the Monster CFT.
The same analysis applied to a second bulk CFT --- a certain $c=2$
Gaussian model --- yields a less strict bound,
suggesting that the precise linear functional bound on $s$ for the Monster CFT is exceptional.
\end{abstract}
\end{titlepage}
\large


\section{Introduction}
Two-dimensional conformal field theories (CFTs) with boundaries have many applications 
to condensed matter systems.  They describe critical boundaries, defects, 
and junctions in 1+1-dimensional quantum critical systems.
In string theory, CFTs with boundary describe the worldsheets of open strings. 

Each conformal boundary condition is characterized by
a number $g$ --- the {\it universal noninteger ground-state 
degeneracy} \cite{AL} ---
defined by
\be
\ln Z(\beta) = c\frac{\pi}{6}\frac{L}{\beta}+\ln g
\ee
in the limit ${L}/\beta \rightarrow \infty$
where
$Z(\beta)={\rm Tr} \, e^{-\beta H} $ is the partition function at inverse temperature $\beta$
of the 1-dimensional system with boundary of length $L$,
and where $c$ is the conformal central charge of the bulk CFT.
The \emph{$g$-theorem} conjectured in \cite{AL} states that
for any renormalization group (RG)
flow between critical boundary conditions for a given fixed
bulk CFT,
the number $g$ is always 
smaller at the IR (final) fixed point than 
at the UV (initial) fixed point.

For arbitrary --- not necessarily critical ---  boundary conditions on a CFT
the boundary entropy $s(\beta)$ is defined in the same fashion,
by subtracting the universal bulk entropy from the 
total entropy $S(\beta)$ in the limit ${L}/\beta \rightarrow \infty$,
\be
S(\beta) = \left (1-\beta\frac\partial{\partial\beta}\right ) \ln 
Z(\beta) = c\frac{\pi}{3}\frac{L}{\beta}+s(\beta)\,.
\ee
The boundary entropy for a critical boundary condition is $s=\ln g$, independent of temperature.
It was shown in \cite{FK} that the boundary entropy $s(\beta)$ for a 
general boundary condition
always decreases with decreasing temperature (contingent on certain 
regularity assumptions
on the ultraviolet properties of the boundary condition).  Equivalently,
$s(\beta)$ decreases under the renormalization group flow.  This proved the $g$-theorem
as a corollary.

In order to control the IR behavior of the renormalization group ---
the low temperature behavior of the boundary system --- it is not 
enough to have a quantity $s(\beta)$ that decreases under the RG
flow.  A lower bound on $s(\beta)$ is needed.  Without a 
lower bound, the RG flow might go on forever towards 
$s(\infty)=-\infty$.  A lower bound on $s(\beta)$ would imply that 
every boundary condition becomes critical at zero temperature.
No way has yet been found to put a lower bound on $s(\beta)$.  

A more modest goal is to establish a lower bound on $s=\ln g$ for all 
the \emph{conformal} boundary conditions for each given bulk CFT.  That 
would at least 
exclude the possibility of flows that end at critical boundary conditions with 
arbitrarily low values of $s$.  Once we have a lower bound on the 
critical values of $s$, it becomes interesting to look for
critical boundary conditions that saturate the bound.  If such 
minimal conformal boundary conditions exist and if any of 
them has a relevant perturbation, the corresponding 
outgoing RG trajectory would have to go on forever, to $s(\infty)=-\infty$.
On the other hand, if such boundary conditions exist
and have no relevant perturbations, it would suggest
that the lower bound on $s$ applies to all boundary conditions, not 
just the critical ones.

In \cite{FKSC} the present authors demonstrated the existence
of a 
lower bound on the boundary 
entropy $s=\ln g$ of all conformal boundary conditions
for any given unitary bulk CFT
subject to the condition that
the lowest scaling dimension $\Delta_{1}$  of the spin-0
bulk fields satisfies $\Delta_{1}>(c-1)/12$.
Only CFTs with $c\ge 1$ were considered because the conformal boundary conditions for the $c<1$ 
unitary CFTs are already completely classified.
No attempt was made in \cite{FKSC} to obtain an optimal lower bound.
The goal was only to show existence of a bound.

Once existence of a lower bound is known, the goal becomes to find the best possible lower 
bound for any given bulk CFT.
Nothing can be assumed about the boundary condition beyond what is 
implied by conformal invariance and the general principles of 
boundary quantum field theory.
On the other hand, all available knowledge of the bulk CFT can be used.
The bound obtained in \cite{FKSC} used only the values of $c$ and 
$\Delta_{1}$.
A sharper lower bound can be obtained for any given bulk CFT by exploiting more 
detailed information about the bulk CFT.
Here,
we carry out this program for two specific bulk CFTs.

Our main example is the Monster CFT.
The Monster CFT is the direct product of the right-moving $c=24$ chiral 
Monster  
CFT with the parity conjugate left-moving $\bar c=24$ chiral Monster CFT.
The chiral Monster CFT was constructed in \cite{FLM}.
Its internal symmetry group is the Monster --- the largest 
finite simple group.
Its fields are all right-moving (holomorphic in 2-dimensional euclidean 
spacetime).
The scaling dimensions of the fields of the chiral theory are 
$2,3,4,\ldots$, so the Monster CFT has spin-$0$ (scalar) fields of dimensions
$\Delta_{k} = 4, 6, 8,\ldots$.
All the spin-0 local couplings are irrelevant,
which makes the Monster CFT of interest for
constructing critical quantum circuits \cite{DF}. 

A variety of conformal boundary conditions (branes) for the Monster 
CFT are known \cite{CGH}.
The lowest value of the boundary entropy among these branes is $s{=}0$ ($g{=}1$).
The known $g{=}1$ branes are the boundary 
conditions that respect the full chiral algebra.
The incoming chiral fields are 
transformed by an element of the Monster symmetry group
and reflected into outgoing chiral fields.
They all have the same boundary spectrum $h_{j} = 2, 3, 4, \cdots$,
with the same multiplicities.

The linear functional (LF) method used in \cite{FKSC} can exploit the 
complete knowledge of the primary spin-$0$ bulk scaling dimensions 
$\Delta_{k}$.
The LF method provides a series of larger and larger numerical 
computations, each of which gives a rigorous lower bound
on $g$.
We find that this series of numerically derived lower bounds on $g$ converges spectacularly 
closely to 1.
We are lead to conjecture that $s\ge 0$ ($g\ge 1$)
is the exact lower bound for the Monster CFT.


Moreover, the LF method can pin down the low-lying spectrum of
boundary scaling dimensions and their multiplicities, for any $g{=}1$
boundary condition, i.e., any boundary condition that saturates the
lower bound.  We show that there are no relevant boundary
perturbations --- that the lowest nonzero boundary scaling dimension
is greater than $1$.  This suggests that the bound $s\ge 0$ holds for
all boundary conditions and that the generic boundary RG flows ends at
a $g{=}1$ conformal boundary condition.
We find strong evidence that the low-lying boundary spectrum for $g{=}1$ branes 
--- including multiplicities --- is uniquely determined,
matching the boundary spectrum of the known $g{=}1$ branes.
This suggests that the known examples comprise all possible $g{=}1$ branes.

The second bulk CFT we study is a certain $c=2$ Gaussian model --- a nonlinear model with a
particular two-torus as target manifold.  Again, we know the complete
spectrum of bulk scaling dimensions and can use the linear functional
method with that knowledge to get a succession of numerical lower
bounds on $g$ that converge rapidly to a limit.  In this case, 
no known brane saturates the lower bound.  Moreover, we show
that no such minimal conformal boundary condition can exist, because
the LF method fixes the multiplicity of the lowest lying boundary
dimension to lie in a range of real numbers that does not include an
integer.  We conclude that the success of the linear functional method
for the Monster CFT is exceptional.

The method is described in sections 2 and 3.
Section 4 presents
the results for the
Monster CFT.
Section 5 presents the results for the $c=2$ Gaussian model.
In section 6 we discuss possible improvements on the linear functional 
method that might give strict lower bounds on $g$ and consistent 
boundary spectra for minimal boundary conditions
for the general bulk CFT.

\section{The linear functional method} 

The linear functional method of producing bounds on quantities in 
conformal field theories was originated in \cite{Cardy:1986}, \cite{Cardy:1991kr} (see Appendix A in particular)  where it was used 
to constrain operator dimensions and state degeneracies in a two-dimensional CFT. 
Similar methods are used in CFTs in higher dimensions,
starting from \cite{RRTV}, and the method was later applied in two dimensions
in particular in \cite{Hellerman:2009bu}. 
Here we briefly summarize the use of the method in \cite{FKSC} to 
get bounds on the boundary entropy. 

\subsection{The modular duality equation}

Consider a given bulk CFT with bulk central charge $c>1$.
A conformal boundary condition is described by a certain bulk state 
$|B\rangle$.
The modular duality equation \cite{Cardy} is 
\be
Z(\beta) =  {\rm Tr} \, e^{-\beta H_{\mathrm {bdry}}} = \langle B| e^{-2\pi H_{\mathrm{bulk}}/\beta} |B\rangle \, . 
\label{eq:modularity}
\ee
Here $H_{\mathrm{bulk}}$ is the hamiltonian for the CFT on a 
circular space, without boundary.
$H_{\mathrm {bdry}}$ is the hamiltonian for the CFT on a line 
interval with the same boundary condition at each end.
The two sides of the duality equation are the two operator 
interpretations of the partition function of the euclidean CFT
on a 2-dimensional annulus.

Expanding both sides of equation \eqref{eq:modularity} in Virasoro characters and 
eliminating the common factor of $1/\eta(i\beta)$ in all the 
characters, we obtain
(see \cite{FKSC} for details, with slightly different notation)
\be\label{eq1}
f_{0}-f_{1} + \sum_{j}N(h_{j}) f_{h_{j}} = g^{2}(\tilde f_{0}-\tilde 
f_{0}) + \sum_{k}b^{2}(\Delta_{k})\tilde f_{\frac12\Delta_{k}}
\ee
where 
\be
f_{h}= \beta^{\frac14}q^{h-\gamma}\, , \quad 
q =e^{-2\pi\beta}\,,\quad
\gamma=\frac{c-1}{24}\,,\quad
\tilde f_{\tilde h}=\tilde\beta^{\frac14}\tilde q^{\tilde h -\gamma}\, ,  \quad
\tilde \beta = \beta^{-1}\,,\quad
\tilde q =e^{-2\pi\tilde \beta}\,,
\ee
and
\begin{itemize}
\item The $\Delta_{k}$ are the distinct scaling dimensions of the primary spin-0 
bulk fields besides the identity, ordered so that
$
0< \Delta_{1}< \Delta_{2}<  
\cdots 
$.
\item The $h_{j}$ are the distinct scaling dimensions of the primary 
boundary fields besides the identity, ordered so that
$
0< h_{1}< h_{2}<  
\cdots 
$.
\item $N(h_{j})$, an integer $\ge 1$, is the 
multiplicity of $h_{j}$.
\item 
$ g= \langle B | 0 \rangle $ is the overlap
between the boundary state and
the bulk ground state $| 0 \rangle $.
\item $
b^{2}(\Delta_{k}) = \sum_{\alpha} \langle B | \alpha \rangle ^{2}
$
where the sum is over all primary spin-0 bulk fields of dimension 
$\Delta_{k}$,
and $|\alpha \rangle$ is the bulk state corresponding to the primary 
field $\phi_{\alpha}$.
\end{itemize}
The numbers $h_{j}$, $N(h_{j})$, $g^{2}$, and $b^{2}(\Delta_{k})$ are properties of the 
boundary condition, so
we assume nothing about them besides the basic constraints from 
unitarity
\begin{itemize}
\item $h_{j}>0$,
\item $N(h_{j}) \text{ integer} \ge 1$,
\item $g^{2}\ge 0$,
\item $b^{2}(\Delta_{k})\ge 0$.
\end{itemize}
On the other hand, we do use the knowledge we have of
the $\Delta_{k}$, which are properties of the bulk CFT.

\subsection{Bounds from linear functionals}
A linear functional $\rho$ acting on functions $F(\beta)$ is a distribution 
\be
\rho( F) = \int_{0}^{\infty}\!\! d\beta\, \hat \rho(\beta)F(\beta) \, . 
\ee
Applying a linear functional $\rho$ to both sides of (\ref{eq1}) we obtain 
\be
\rho(f_{0}-f_{1}) + \sum_{j}N(h_{j})\rho(f_{h_{j}}) = 
g^{2}\rho( \tilde f_{0}-\tilde f_{1}) + \sum_{k}b^{2}(\Delta_{k})\rho( \tilde 
f_{\frac12\Delta_{k}})
\label{eq:modrho}
\ee
If we can choose $\rho$ so that 
\be\label{cond1}
\rho( f_{h})\ge 0  \quad \forall h>0\, , 
\ee
\be\label{cond2}
\rho( - \tilde f_{\frac12\Delta_{k}})\ge 0 \quad \forall \Delta_{k}
\ee
we get an inequality 
\be
g^{2}\rho( \tilde f_{0}-\tilde f_{1}) \ge \rho(f_{0}-f_{1})\, . 
\label{eq:lowerbound1}
\ee
It was shown in \cite{FKSC} that condition \eqref{cond1} implies $\rho( \tilde 
f_{0}-\tilde f_{1})>0$, so the inequality is a lower bound on $g^{2}$.
Equations \eqref{eq:modrho}--\eqref{eq:lowerbound1} are indifferent to positive rescalings of $\rho$, so we might as well 
impose the normalization condition
\be
\rho( \tilde f_{0}-\tilde f_{1}) =1\,.
\label{eq:norm1}
\ee
Equation \eqref{eq:modrho} becomes
\be
g^{2} =\rho(f_{0}-f_{1})
+ \sum_{j}N(h_{j})\rho(f_{h_{j}}) 
+ \sum_{k}b^{2}(\Delta_{k})\rho( -\tilde f_{\frac12\Delta_{k}})
\label{eq:grho}
\ee
and the lower bound is
\be
g^{2} \ge \rho( f_{0}-f_{1})\, . 
\label{eq:bound1}
\ee
Maximizing over all distributions $\rho$ subject to the positivity 
conditions \eqref{cond1} and 
\eqref{cond2} and the normalization condition \eqref{eq:norm1},
we obtain the optimal linear functional bound
\be
g^{2}\ge g_{B}^{2}= {\rm max}_{\rho} \;\rho( f_{0}-f_{1})\,.
\ee
The lower bound $g_{B}^{2}$ depends on the bulk central charge $c$ and on the entire bulk spin-0 spectrum 
$\Delta_{k}$.
In \cite{FKSC}, the goal was to show the existence of a lower bound 
on $g$ for as general a class of bulk CFTs as possible,
so condition (\ref{cond2}) 
was replaced by the stronger condition
\be
\rho( - \tilde f_{\frac12\Delta})\ge 0\, , \quad \forall \Delta \ge \Delta_{1} \,,
\label{eq:cond2prime}
\ee
which gives a lower bound that depends only on $c$ and $\Delta_{1}$.
It was shown that
conditions \eqref{cond1} and \eqref{eq:cond2prime} 
can be satisfied together if and only if $\Delta_{1}> 2\gamma= 
(c-1)/12$.

\subsection{Practical calculations}
In practice, the method is to maximize over larger and larger finite 
dimensional subspaces of linear functionals
of the form
\be
\rho( F) = \mathcal{D} F(\beta)\,,
\label{eq:diffop1}
\ee
where $\mathcal{D}$ is a polynomial differential operator 
in $\beta$ of order $2n-1$, and  $\mathcal{D}F$ is evaluated at some
fixed value of $\beta$.
The order of the differential operator must be odd because of
the positivity conditions.
For each $n$, we maximize over the $2n$-dimensional space of differential operators 
of order $2n-1$.

It is impractical to enforce the positivity condition~\eqref{cond2} for the 
infinite collection of $\Delta_{k}$.  Instead, we enforce the 
stronger condition
\be
\rho(- \tilde f_{\frac12\Delta})\ge 0 \quad \text{for }
\Delta = \Delta_{1}, \,\Delta_{2},\,\ldots,\Delta_{N-1},
\text{ and for } \Delta\ge \Delta_{N}\,.
\label{cond2pract}
\ee
For each value of $n$ and $N$, we
get a lower bound on $g$.
As we increase $n$ or $N$, the lower bound gets larger.
The limit $N\rightarrow\infty$ will realize the positivity 
condition~\eqref{cond2}.
The limit $n\rightarrow\infty$ will exhaust the space of 
linear functionals
because any linear functional on real analytic 
functions can be approximated by a differential operator 
$\mathcal{D}$ acting at a single point $\beta$.
The combined limit $N,n\rightarrow\infty$ gives the optimal LF bound.

In practice,
we solve the maximization problem numerically for various values of 
the parameters $n,N$, limited by computational resources.
We use a more or less arbitrary value of $\beta$.
The numerical solution of each maximization problem is of course not
an exact solution.
The numerical solution does provide a concrete 
linear functional $\rho_{n,N}$.
We verify that $\rho_{n,N}$ satisfies the positivity conditions.
Then we calculate
a rigorous lower bound on $g^{2}$
using equation \eqref{eq:lowerbound1}.
Thus each numerical maximization provides a rigorous lower bound on 
$g^{2}$.

\subsection{Integrality constraints}
Note that the linear functional method makes no use of the fact that 
the boundary multiplicities $N(h_{j})$ must be integers $\ge 0$.
The linear functional bound is a necessary condition for the existence of 
a solution to equation \eqref{eq1} with real $N(h_{j})> 0$,
which of course is also a necessary condition for a solution with integer 
$N(h_{j})>0$.
We comment in the final section on the possibility of finding 
better lower bounds on $g$ that take account of the integrality 
constraints.

\subsection{The existence of solutions to the modular duality equation}
The optimal linear functional bound is a necessary 
\emph{and} sufficient condition for existence of a solution to 
equation \eqref{eq1} with $N(h_{j})$ real,
by a small variation of an argument used in \cite{extr_func}.
Let $\mathcal{F}$ be the space of functions of $\beta>0$ (suitably 
defined).  
For any real non-negative measure $N(h)$ on $h>0$, and any collection 
of numbers $b^{2}(\Delta_{k})\ge 0$ define $f[N,b^{2}]\in \mathcal{F}$
by
\be
f[N,b^{2}] = \int_0^{\infty} dh\, N(h) f_{h}+
\sum_{k} b^{2}(\Delta_{k}) 
\left (-\tilde f_{\frac12 \Delta}\right)
\,.
\ee
The functions $f[N,b^{2}]$ form
a convex cone $C$ in $\mathcal{F}$
\be
C = \bigg\{f[N,b^{2}]\bigg  \}.
\ee
Define a vector $v\ne 0$ in $\mathcal{F}$
\be
v = g^{2}(\tilde f_{0}-\tilde f_{1}) - (f_{0}-f_{1})\,.
\ee
There exists a real solution of equation \eqref{eq1} iff $v \in C$.
The \emph{Generalized Farkas Lemma} \cite{Craven:1977} 
says
\begin{quote}
$v \in C$ iff there is no hyperplane
separating $v$ from C or, equivalently, iff $\rho(v) > 0$ for all linear functionals $\rho$ 
satisfying $\rho(C) \ge 0$.
\end{quote}
The condition $\rho(C) \ge 0$ is exactly conditions \eqref{cond1} and 
\eqref{cond2}.
The condition $\rho(v) > 0$ is exactly equation \eqref{eq:bound1}.
Therefore there exists a real solution of equation \eqref{eq1} iff 
$g^{2}$ satisfies the optimal linear function bound.

\section{Map to an SDP problem}

We next recast the maximization problem as
a semidefinite programming (SDP) problem, following \cite{RRTV,PSD}. 
An SDP problem is an optimization over a set of positive-semidefinite 
matrices --- the SDP variables.
The problem is to minimize an \emph{objective function} $\mathcal{O}$ 
which is linear in the SDP variables, subject to a collection of 
equality constraints also linear in the SDP variables.
Effective codes are available for solving SDP problems 
numerically.

The general differential operator $\mathcal{D}$ of equation 
\eqref{eq:diffop1} can be written ${\cal D}=D(-4\beta\partial_{\beta})$ 
for $D(z)$ a polynomial of degree $2n-1$.
Recall that $\mathcal{D}$ is acting at a specific fixed value of $\beta$.
Maximizing over differential operators $\mathcal{D}$ is equivalent to 
maximizing over polynomials $D(z)$.
Now define two polynomials $p(x)$ and $\tilde p(x)$, each of degree 
$2n-1$, by
\be
p(x) = x^{-\frac{1}{4}}e^{x/4} D(-4x\partial_{x})\left( x^{\frac{1}{4}}e^{-x/4}\right) \, , \quad 
\tilde p(x) = -x^{-\frac{1}{4}}e^{x/4} D(4x\partial_{x})\left( x^{\frac{1}{4}}e^{-x/4}\right) \, . 
\label{eq:pptfromD}
\ee
We will see shortly that the map from polynomials $D(z)$ to 
polynomials $p(x)$ is invertible, as is the map from $D(z)$ to 
$\tilde p(x)$.  Thus we can maximize over polynomials $p(x)$, or over 
polynomials $\tilde p(x)$.
Actually, we will maximize over pairs of polynomials $p(x)$, $\tilde 
p(x)$ subject to the constraint that they come from the same differential 
operator $\mathcal{D}$.

The definitions of $p(x)$ and $\tilde p(x)$ were designed so that
\ateq{3}{
\mathcal{D} f_{h} &= P(h) f_{h}\,,\qquad
&\text{ where }\, P(h) &= p(x(h))\,, \quad
&x(h) &= 8\pi\beta (h-\gamma)\,,\\
\mathcal{D} \tilde f_{\tilde h} &= - \tilde P  (\tilde h)
f_{\tilde h}\,,\quad
&\text{ where }\, \tilde P (\tilde h) &= \tilde p(\tilde x(\tilde h)) \,,
&\tilde x(\tilde h) &= 8\pi \tilde \beta (\tilde h - \gamma)\,,
}
so the positivity conditions \eqref{cond1} and \eqref{cond2pract} on the 
differential operator $\mathcal{D}$ are equivalent to
positivity conditions on the polynomials $p(x)$ and $\tilde p(x)$,
\ateq{2}{
p(x) &\ge 0 \quad &&\text{for } x\ge x(0) \label{cond1a}
\\
\tilde p (\tilde x) &\ge 0 \quad &&
\text{for }\tilde x = \tilde x_{1},
\ldots, \tilde x_{N-1}
\,\text{ and }\, \tilde x\ge \tilde x_{N}\,,
\;\text{ where }\tilde x_{k} = \tilde x\left({\textstyle\frac12} 
\Delta_{k}\right )\,.
}
Equation \eqref{eq:modrho} --- which is $\mathcal{D}$ applied to both 
sides of the modular duality equation \eqref{eq1} --- now reads
\be
P(0) f_{0}-P(1)  f_{1} +\sum_{j}N(h_{j})P(h_{j})f_{h_{j}} 
= 
g^{2}[\tilde P(1)\tilde f_{1}- \tilde P(0) \tilde f_{0}] 
- \sum_{k}b^{2}(\Delta_{k}) \tilde P\left ({\textstyle\frac12} 
\Delta_{k}\right ) \tilde f_{\frac12\Delta_{k}}\, .
\label{eq:Dmod}
\ee
The normalization condition \eqref{eq:norm1} becomes
\be
- \tilde P(0) \tilde f_{0}+\tilde P(1)\tilde f_{1} = 1\,,
\ee
giving
\be
\label{eq:gformula}
g^{2} =
g^{2}_{B}[p,\tilde p]  + \sum_{j}N(h_{j})P(h_{j})f_{h_{j}} 
+ \sum_{k}b^{2}(\Delta_{k}) \tilde P\left ({\textstyle\frac12} 
\Delta_{k}\right ) \tilde f_{\frac12\Delta_{k}}
\ee
where
\be
g^{2}_{B}[p,\tilde p] = P(0) f_{0}-P(1)  f_{1} 
\label{eq:objective}
\ee
is the lower bound to be maximized over pairs of polynomials $p(x)$, $\tilde 
p(x)$ to get the optimal bound for each $n,N$,
\be
g^{2}_{n,N} = max_{p,\tilde p}\; [P(0) f_{0}-P(1)  f_{1} ]\,.
\ee

Again following~\cite{RRTV}, 
we write the general solution of
the continuum positivity constraints on
the polynomials $p(x)$ and $\tilde p(\tilde x)$ in 
terms of positive semidefinite $n{\times} n$ matrices $Y_{\alpha}$ \cite{PSD},
\aeq{
p(x) &= \sum_{k=0}^{2n-1}p_{k}x^{k} = \mathbf{x}^{t} Y_{1} \mathbf{x} + (x-x(0)) \mathbf{x}^{t} 
Y_{2} \mathbf{x} \label{eq:polysY}\\
\tilde p(x) &= \sum_{k=0}^{2n-1}\tilde p_{k}x^{k} = \mathbf{x}^{t} Y_{3} \mathbf{x} + (x-\tilde x_{N}) \mathbf{x}^{t} 
Y_{4} \mathbf{x} \,.\nonumber
}
where $\mathbf{x}$ is the $n$-vector with components
$(1,x,x^{2},\ldots,x^{n-1})$.
Note that the polynomial coefficients $p_{k}$ and $\tilde p_{k}$ are linear 
functions of the matrix elements of the $Y_{\alpha}$.
The remaining positivity constraints are
\be
\tilde p(\tilde x_{k}) \ge 0\,,\quad k=1,\ldots,N-1 \,.
\label{eq:inequalities}
\ee
These are solved by introducing $N-1$ auxiliary $1{\times} 1$
positive semidefinite matrices $y_{k}$ subject to the $N-1$ equality 
constraints
\be
y_{k} = \tilde p(\tilde x_{k})\,,  \quad k=1,\ldots,N-1 \,.
\label{eq:constr2}
\ee

Finally, we need to impose the condition that the polynomials 
$p(x)$ and $\tilde p(\tilde x)$ come from the same differential 
operator $\mathcal{D}=D(-4\beta\partial/\partial\beta)$.
The differential operator ${\cal D}=D(-4\beta\partial_{\beta})$,
\be
D(z)=\sum_{l=0}^{2n-1} d_{l} z^{l}\,,
\label{eq:Dexpansion}
\ee
is determined by the 
coefficients of either of the two polynomials by the equations
\be
d_{l} = \sum_{k\ge l} g_{lk} p_{k} = - \sum_{k\ge l} 
(-1)^{l}g_{lk}\tilde p_{k}\
\label{eq:constraints}
\ee
where the numbers $g_{lk}$ --- which depend only on $n$ ---- are 
calculated in Appendix~\ref{app:glk}.
Therefore the condition that the two polynomials 
come from the same differential 
operator is expressed by $2n$ equality constraints
\be \label{eq_constr1}
\sum_{k\ge l} [g_{lk} p_{k} + (-1)^{l}g_{lk}\tilde p_{k}] = 0 \, , \enspace \enspace l=0,1,\dots ,2n-1
\ee 
which are linear constraints on the matrix elements of the semidefinite matrices.

The maximization problem is now re-formulated as an 
SDP problem:
\begin{itemize}
\item The SDP variables are the semidefinite  matrices $Y_{\alpha}$, $\alpha = 1,\ldots,4$
and  $y_{k}$, $k=1,\ldots,N-1$.
The $Y_{\alpha}$ are $n{\times} n$ matrices.  The $y_{k}$ are
$1{\times} 1$ matrices.
\item There are $2n$ equality constraints given by equation \eqref{eq_constr1}
and $N-1$ equality constraints given by equation \eqref{eq:constr2}.
\item The objective function to be maximized is $\mathcal{O} = g^{2}_{B}[p,\tilde p]$ given 
by equation \eqref{eq:objective}.
\end{itemize}
The equality constraints and the objective function are
all linear functions of the matrix elements of the semidefinite 
matrices.

Following the lead of \cite{PSD}, 
we used the arbitrary precision SDP 
solver SPDA-GMP \cite{SDPA}, which calculates using the GMP arbitrary precision arithmetic 
libraries.
We found it necessary to calculate using extended precision floating point arithmetic
in order to obtain stable numerical solutions to the SDP problems.
In practice, we found it useful and feasible to solve our
SDP problems with 400 decimal digits of precision.

We prepare the SPDA-GMP problem specifications in the Sage symbolic 
mathematics program \cite{SAGE}.
The input for each run consists of
\begin{itemize}
\item the central charge $c$ and the list of the low-lying $\Delta_{k}$, $k=1,\ldots,N$ in the bulk CFT,
\item the integer $n$ specifying the rank of each of the semidefinite 
matrices $Y_{\alpha}$ and the order $2n-1$ of the differential 
operator.
\end{itemize}
We scan increasing values of $n$ and $N$ to the limits of our
computational resources.

\subsection{Verification of numerical solutions}

For each choice of $n$ and $N$, the SDP solver returns a set of 
semidefinite matrices $Y_{\alpha}$ that solves the 
optimization problem approximately.  The solver is a black box to us, so we 
cannot take the solution at face value.  We verify that the SDP 
solution actually provides a rigorous lower bound on $g^{2}$.

From the solution 
matrices $Y_{\alpha}$ provided by the solver, we calculate the polynomials $p(x)$ and $\tilde 
p(x)$ by equation~\eqref{eq:polysY}.  From each of the two polynomials, we calculate the 
coefficients of the corresponding 
differential operator $\mathcal{D}$.  This gives us two slightly 
different differential 
operators, because the solver does not impose the equality constraints 
exactly.  Then we reverse the 
calculation for each of the two differential operators.  From the differential operator we 
calculate the corresponding polynomials $p(x)$ and $\tilde p(x)$
and check that they satisfy the 
positivity constraints.
We check $\tilde p(x_{k})\ge 0$ for $k=1,\ldots,N$ by direct 
calculation.
We check positivity in the half-line,
$p(x)\ge 0$ for $x\ge x_{1}$ and $\tilde p (x) \ge 0$ for $x\ge 
\tilde x_{N}$, 
in two ways.
First, we find all real roots of $p(x)$ numerically (in 
Sage) to check that all are less than than $x_{1}$.
We do the analogous check for $\tilde p(x)$.
Second, we check that the absolute minimum of $p(x)$ for $x\ge 
x_{1}$ is nonnegative by
finding all real roots of $p'(x)$ with $x\ge x_{1}$ and 
then finding the minimum value of $p(x)$ at those roots of
$p'(x)$.  We do the analogous check for 
$\tilde p(x)$.

Sometimes the positivity checks fail,
presumably
because the SDP solver enforces the positivity constraints with too 
much tolerance.
When the positivity constraints are satisfied for at least one of the 
two differential operators reconstructed from the $Y_{\alpha}$,
the resulting lower bound on $g^{2}$ is rigorous, since it
derives from 
a specific linear functional given as a specific differential 
operator acting at a specific value of $\beta$.
The calculated lower bound is not the best 
possible bound for the given values of $n$ and $N$, but it is a rigorous bound.
Strictly speaking, we should control the rounding errors by using rigorous interval arithmetic
in the calculations to check the validity of the solutions.
We do not go to such lengths.
Instead,
we do the numerical calculations with a floating point precision of 400 decimal digits,
which is far more than enough to allow us to disregard rounding errors.
We have checked that
Sage calculates the roots of polynomials accurately to within a few digits of the 
floating point precision,
and that 
the positivity checks are passed by tolerances which are hundreds of 
orders of magnitude larger than our floating point precision.


\section{Numerical results for the Monster CFT} 
The $c=24$ chiral Monster CFT \cite{FLM} 
is the algebra of right-moving (holomorphic) 
fields ${\cal H}_{\mathrm {M}}$
constructed as the chiral ${\mathbb Z}_{2}$ orbifold
of the holomorphic vertex operator algebra associated to the 
24-dimensional self-dual Leech lattice (see e.g.\ \cite{Gannon} or 
\cite{CGH} section 2 for details of the construction).
The Monster group --- the largest finite simple group ---
is the internal symmetry group of the chiral Monster CFT.
Each element $\gamma$ of the Monster group acts on the fields of the 
chiral Monster CFT by $\phi(z) \mapsto \phi^{\gamma}(z)$.
All we use from this construction is the spectrum $h_{k} = 
2,3,4,\ldots$ of distinct non-zero primary conformal weights.

The Monster CFT is made by tensoring together 
the right-moving  chiral Monster CFT
with its left-moving conjugate
$
{\cal H}_{\mathrm {bulk}} = {\cal H}_{\mathrm {M}}\otimes \bar {\cal H}_{\mathrm {M}} \, .
$
Each primary field of the Monster CFT has the form $\phi(z,\bar z) 
= \phi_{R}(z)\bar \phi_{L}(\bar z)$, with scaling dimension $\Delta=h+\bar 
h$ and spin $h-\bar h$, where $h$ and $\bar h$ are the conformal 
weights of the chiral primaries $\phi_{R}$ and $\phi_{L}$.
So
the distinct scaling dimensions of the spin-0 primary fields 
are
\be
\Delta_{k} = 2k + 2 \, , \enspace k\ge 1 \ .  
\ee
We took $\beta= 0.93$ for all our Monster calculations.
This is approximately the value of $\beta$ which maximizes the $n=1$, $N=1$
lower bound on $g$.

The Monster branes \cite{CGH} with the smallest known value 
$g{=}1$ are in one-to-one correspondence with
the elements $\gamma$ of the Monster group.
The brane is
given by a bulk state $\langle \gamma|$
on the unit circle $|z|=1$
satisfying
\be
\langle \gamma| \, \phi(z)(dz)^{h} = \langle \gamma| \, 
\bar\phi^{\gamma}(1/\bar z)(d(1/\bar z))^{h}
\ee
for all the primary chiral fields $\phi(z)$.  
For each of these branes, the partition function of the unit interval 
is the same
\be
Z = J(i\beta) = j(i\beta) -744 = 
q^{-1} + 196884 q + 21493760 q^{2} + 864299970 q^{3}+ O(q^4)
\ee
where $j(\tau)$ is the $j$-invariant.
Since $j(\tau) = j(-1/\tau)$, each of these branes has $g{=}1$.  To 
find the boundary primary dimensions $h_{j}$ and their multiplicities 
$N(h_{j})$, we expand the partition function in the $c=24$ Virasoro characters
\be
Z = \frac{(1-q)q^{-1} + \sum_{j} N(h_{j}) q^{h_{j}-1}}{\prod_{n=1}(1-q^{n})}
\ee
to get
\be
\sum_{j} N(h_{j}) q^{h_{j}-1} = 196883 q + 21296876 q^2 + 842609326 q^3 + O(q^4)
\,.
\label{eq:g=1mult}
\ee
So the spectrum is $h_{j}= 2,3,4,\ldots$ with multiplicities 
$N(2)=196883$, $N(3)= 21296876$, $N(4)=842609326$, \ldots.

\subsection{Lower bounds on $g$}
Let us write the rigorous lower bounds in the form
\be
g^{2}\ge g^{2}_{n,N} = 1 - \epsilon_{n,N}
\,.
\ee
The following table gives the values of
$\epsilon_{n,N}$ we found for SDP solutions that passed the positivity tests:
\vspace{-2ex}

\be
\label{Monsterlowerbounds}
\begin{array}{cc|c|c|c|}
\multicolumn{2}{c}{} &\multicolumn{3}{c}{N} \\
\multicolumn{2}{c}{}& \multicolumn{1}{c}{15} & \multicolumn{1}{c}{31} & \multicolumn{1}{c}{41} \\
\cline{3-5}
\multirow{5}{*}{n} &15 &1.93{\times} 10^{-4} & &  \\
&24 & & &   1.08{\times} 10^{-9} \\
&30 & &1.12 {\times} 10^{-10} & 6.23{\times} 10^{-11}   \\
&36 & & 7.25{\times} 10^{-13} &   \\
&42 & &6.03{\times} 10^{-19} &  \\
\cline{3-5}
\end{array}
\ee
\vskip3ex\noindent
Our best lower bound is
\be\label{num_bound}
g^{2}\ge g_{42,31}^{2} =  1 - 6.03 {\times} 10^{-19}
\ee 
which gives a lower bound on the boundary entropy
\be
s \ge -3.02 {\times} 10^{-19}\,.
\ee
This is a rigorous bound, since it is derived from a specific 
differential operator that satisfies the positivity constraints.

It seems reasonably clear that the lower bounds shown in table 
\ref{Monsterlowerbounds} are converging to the optimal bound $g^{2}\ge 1$,
$ s \ge 0$.
Given that the smallest value of $g$ for the known conformal boundary conditions 
\cite{CGH} is $g= 1$, our numerically derived bounds give very strong indication 
that $g\ge 1$ is the exact lower bound for all possible Monster branes.

\subsection{The boundary spectrum for minimal branes}
\label{subsect:minimalbranes}

Let us call a conformal boundary condition \emph{minimal} if it 
saturates the optimal linear functional lower bound,
$g=g_{B}$.
Equation \eqref{eq:gformula} for an optimal linear functional is
\be
g^{2} =
g^{2}_{B}  + \sum_{j}N(h_{j})P_{\mathit{opt}}(h_{j})f_{h_{j}} 
+ \sum_{k}b^{2}(\Delta_{k}) \tilde P_{\mathit{opt}}\left ({\textstyle\frac12} 
\Delta_{k}\right ) \tilde f_{\frac12\Delta_{k}}
\ee
where $P_{\mathit{opt}}$ and $\tilde P_{\mathit{opt}}$ are 
nonnegative.
Thus $g=g_{B}$ requires the $h_{j}$ to lie at 
zeros of the function $P_{\mathit{opt}}(h)f_{h}$.
(It also follows that the $b^{2}(\Delta_{k})$ can be non-vanishing only
when $\Delta_{k}/2$ is at a zero
of $\tilde P_{\mathit{opt}}(\tilde h) \tilde f_{\tilde h}$, but we do not pursue this point.)

For each of the solutions returned by the SDP solver,
we calculated the local minima of the function $P(h)f_{h}$ for $h\ge  
0$.  As $n$ and $N$ increase, successively more of the local minima 
approach the values $2,3,4,\ldots$ and the values of $P(h)f_{h}$ at those 
local minima approach zero.
For our best solution, with $n=42$, $N=31$,
the first 8 local minima are presented in the table below.
\be
\begin{array}{|c|c|c|}
\hline
h & P(h) & P(h) f_{h} \\
\hline
2 -1.014107{\times} 10^{ -17}& 2.547717 {\times} 10^{-23}&5.685816{\times} 10^{-26}\\
3 +3.532221{\times} 10^{-16} & 8.126899 {\times} 10^{-23}&5.258071{\times} 10^{-28}\\
4 -3.099596{\times} 10^{-15} & 7.080475 {\times} 10^{-22}&1.328079{\times} 10^{-29}\\
5 +2.776280{\times} 10^{-14} & 1.063502 {\times} 10^{-20}&5.783089{\times} 10^{-31}\\
6 +1.070319{\times} 10^{-14} & 2.274723 {\times} 10^{-19}&3.585999{\times} 10^{-32}\\
7 -1.168940{\times} 10^{-12} & 6.281304 {\times} 10^{-18}&2.870723{\times} 10^{-33}\\
8 -2.778270{\times} 10^{-11} & 2.108470 {\times} 10^{-16}&2.793631{\times} 10^{-34}\\
9 -3.057268{\times} 10^{-10} & 8.266698 {\times} 10^{-15}&3.175364{\times} 10^{-35}\\
\hline
\end{array}
\ee
The evidence seems reasonably strong that $P_{\mathit{opt}}(h)f_{h}$ will 
have zeros for $h>0$ exactly at $h=2,3,4,\cdots$,
so any $g{=}1$ brane must have boundary spectrum $h_{j}=2,3,4,\cdots$.

\subsection{Stability of $g{=}1$ branes}
Our numerical results prove that any $g{=}1$ brane must be 
stable, i.e.\ that the lowest boundary scaling dimension of a $g{=}1$ 
brane satisfies $h_{1}>1$.

For the $n=42$, $N=31$ solution, equation~\eqref{eq:gformula} becomes
\be
(g^{2}-1) + 6.03 {\times} 10^{-19} =  \sum_{j}N(h_{j})P(h_{j})f_{h_{j}} 
+ \sum_{k}b^{2}(\Delta_{k}) \tilde P\left ({\textstyle\frac12} 
\Delta_{k}\right ) \tilde f_{\frac12\Delta_{k}}
\ee
When $g=1$, this implies
\be
6.03 {\times} 10^{-19} > P(h_{1})f_{h_{1}} 
\ee
since $N(h_{1})\ge 1$.
We check that $ P(h)f_{h}>6.03 {\times} 10^{-19} $
for $0<h\le 2- 2.1 {\times} 10^{-7}$, so we conclude that
\be
h_{1}>  2- 2.1 {{\times}} 10^{-7}
\,.
\ee
Therefore any $g{=}1$ brane must be stable.

\subsection{Boundary multiplicities}
\label{subsect:boundarymults}

A small modification of the SDP problem gives upper and 
lower bounds on 
the boundary multiplicities $N(h_{j})$ for any $g{=}1$ brane with 
boundary spectrum $h_{j}=2,3,4,\ldots$.
The multiplicities must be integers, so sufficiently tight bounds fix 
them precisely.  We find $N(2) =196883$,  $N(3)=21296876$, $N(4)=842609326$
in exact agreement with the boundary multiplicities of the known 
$g{=}1$ branes as given by equation~\eqref{eq:g=1mult}.

When $g=1$, equation~\eqref{eq:Dmod} becomes
\be
\sum_{j}N(h_{j})P(h_{j})f_{h_{j}} 
+ \sum_{k}b^{2}(\Delta_{k}) \tilde P\left ({\textstyle\frac12} 
\Delta_{k}\right ) \tilde f_{\frac12\Delta_{k}}
=
-[\tilde P(0) \tilde f_{0}-\tilde P(1)\tilde f_{1} + P(0) f_{0}-P(1)  f_{1}]
\label{eq:Dmodg=1}
\ee
which gives an inequality
\be
N(h_{1}) P(h_{1})f_{h_{1}} \le 
-[\tilde P(0) \tilde f_{0}-\tilde P(1)\tilde f_{1} + P(0) f_{0}-P(1)  f_{1}]
\ee
if we enforce the positivity conditions
\ateq{2}{
P(h_{j}) &\ge 0 \quad &\text{ for } j&=2,3,\ldots \\
\tilde P\left ({\textstyle\frac12} 
\Delta_{k}\right )  &\ge 0 \quad &\text{ for } k&=1,2,3,\ldots\,.
}
Using the normalization condition $P(h_{1})f_{h_{1}}=1$ we get an upper 
bound
\be
N(h_{1}) \le -[\tilde P(0) \tilde f_{0}-\tilde P(1)\tilde f_{1} + P(0) f_{0}-P(1)  f_{1}]\,,
\ee
while using the normalization $P(h_{1})f_{h_{1}}=-1$ gives a lower bound
\be
\tilde P(0) \tilde f_{0}-\tilde P(1)\tilde f_{1} + P(0) f_{0}-P(1)  f_{1} \le N(h_{1})\,.
\ee
For both normalizations, we want to maximize the objective function
\be
\mathcal{O}= 
\tilde P(0) \tilde f_{0}-\tilde P(1)\tilde f_{1} + P(0) f_{0}-P(1)  f_{1}
\ee
to get the optimal bounds on $N(h_{1})$.

For computability, as before, we replace the infinite series of positivity conditions 
with the stronger conditions
\ateq{2}{
P(h) &\ge 0 \quad &\text{ for }h&=h_{j},\; j=2,3,\ldots,N-1 \text{ and } 
h\ge h_{N} \\
\tilde P (\tilde h  )  &\ge 0  &\text{ for }
\tilde h &= {\textstyle\frac12} \Delta_{k},\;
k=1,2,\ldots,N-1
 \text{ and } 
\tilde h\ge {\textstyle\frac12} \Delta_{N}
\,.
}
Some numerical results are given in the table below.
\be
\begin{array}{|r|r|r@{\,<\delta<\,}l|}
\hline
n & N & \multicolumn{2}{|c|}{\text{bounds on }\delta = N(2)- 196883}\\
\hline
8 & 5 & -0.79 & 0.74\\
10 & 10 & -2.7{\times}10^{-5} &4.8{\times}10^{-4}\\
12 & 8 & -3.6{\times}10^{-6} & 8.3{\times}10^{-5}\\
12 & 10 & -2.9{\times}10^{-7} & 6.2{\times}10^{-6}\\
15 & 20 &  -1.3{\times}10^{-10}& 5.5{\times}10^{-12}\\
\hline
\end{array}
\ee
The $n=8$, $N=5$ bounds are enough to fix $N(2)= 196883$,
since the multiplicities $N(h_{j})$ must be integers.
The additional results illustrate convergence to a 
sharp optimal bound.
These 
calculations were done at $\beta=1.0$.

Now we can substitute $N(2)=196883$ into equation~\eqref{eq:Dmodg=1} to 
get bounds on $N(3)$.
\be
\begin{array}{|r|r|r@{\,<\delta<\,}l|}
\hline
n & N & \multicolumn{2}{|c|}{\text{bounds on }\delta = N(3)- 21296876}\\
\hline
8 & 5 & -38 & 61\\
10 & 5 & -2.6 &4.4\\
10 & 8 & -8.5{\times}10^{-3} & 8.1{\times}10^{-3}\\
\hline
\end{array}
\ee
So we have $N(3)=21296876$ and can calculate bounds on $N(4)$.
\be
\begin{array}{|r|r|r@{\,<\delta<\,}l|}
\hline
n & N & \multicolumn{2}{|c|}{\text{bounds on }\delta = N(4)- 842609326}\\
\hline
10 & 10 & -8.3{\times}10^{-2} & 2.2{\times}10^{-2} \\
10 & 12 & -8.3{\times}10^{-2} &3.6{\times}10^{-3}\\
15 & 15 & -1.2{\times}10^{-9} &7.0{\times}10^{-9}\\
\hline
\end{array}
\ee
So $N(4)=842609326$.  

At this point we extrapolate to the 
conclusion that any $g{=}1$ brane must have the same spectrum 
and multiplicities as the known $g{=}1$ branes.

\section{$c=2$ Gaussian model}

Our second example is a certain $c=2$ Gaussian model --- a nonlinear model 
whose target space is a 2-torus
whose radii are  both equal to 
$R=\sqrt{2}R_{\mathrm{sd}}$ where 
$R_{\mathrm{sd}}$ is the self-dual radius.
There is no $B$-field in this example. 
All known conformal boundary conditions for this CFT have 
$g\ge 0.5$ \cite{Affleck_etal}.

We show in Appendix~\ref{app:c=2} that the spin-0 scaling dimensions 
of the Virasoro primary fields are
\be
\left\{ \Delta_{k} \right\} = \left \{ 
\frac{m}4 \;: m>0, m \equiv 0,1,2,4,5\, (\mathrm{mod}\,8)
\right \}
\ee
Using this list of scaling dimensions,
we calculated lower bounds on $g^{2}$ as before,
only with a different value of $c$ and a different list of $\Delta_{k}$.
For $n=36$, $N=40$ (using $\beta=1.0$)
we obtained the lower bound 
\be
g^{2} > 0.1009\,.
\ee
The bound did not improve appreciably when we increased $n$ from $24$ 
to $30$ and then to $36$.
This linear functional bound is
well below the smallest known value $g^{2}=0.25$.

We next explored the possibility of a minimal brane, that saturates the 
linear functional bound.
For a minimal brane,
the boundary scaling dimensions $h_{j}$ must lie among the zeros of
the function $P_{\mathit{opt}}(h)f_{h}$,
as in section~\ref{subsect:minimalbranes}.
For the $n=36$, $N=40$ solution, the first ten local minima of the function $P(h)f_{h}$
are shown in the following table.
\be
\begin{array}{|c|c|}
\hline
h & P(h)f_{h} \\
\hline
2.527099 &  1.801512{{\times}} 10^{-64}\\
4.281833 &  9.260807{{\times}} 10^{-65}\\
5.802231 &  5.951241{{\times}} 10^{-65}\\
7.160648 &   3.699803{{\times}} 10^{-65}\\
8.443321 &   2.115923{{\times}} 10^{-65}\\
9.768486 &   1.235336{{\times}} 10^{-65}\\
11.03488 &   8.684430{{\times}} 10^{-66}\\
12.33631 &   5.141452{{\times}} 10^{-66} \\
13.67932 &   3.325586{{\times}} 10^{-66} \\
15.06931 &   2.052798{{\times}} 10^{-66} \\
  \hline
  \end{array}
\ee
The low-lying $h_{j}$ should be from this list.

With these values for the low-lying $h_{j}$, we determined the 
boundary multiplicities as in section~\ref{subsect:boundarymults}.
Using $n=10$, $\beta=1.0$, and taking account of the first $N=24$ of 
the bulk $\Delta_{k}$, we obtained the bounds
\be
6.30974 < N(h_{1}) < 6.30978\,. 
\ee
But $N(h_{1})$ is an integer.
Therefore the linear functional bound on $g$ cannot be saturated.
The true lower bound on $g$ must be higher than the linear 
functional bound.

\section{Conclusions}
We have proved by numerical computation (1) a lower bound on the boundary entropy $s$ of 
a Monster brane, (2) the stability of branes saturating the bound, and 
(3) the uniqueness of the low-lying boundary spectrum of such 
extremal branes.
Our numerical results give 
strong evidence for the exact $s{=}0$ ($g{=}1$) lower bound on the boundary entropy of 
Monster branes
and for the uniqueness of the boundary spectrum of such extremal branes.
The lower bound on $s$ and the stability of the extremal branes suggests 
that $s\ge 0$ for all boundary conditions, conformal or not.

The $c=2$ example shows that this situation is exceptional, that in general the 
optimal linear function bound may not be the true lower bound. It would be interesting to have some clues 
as to when the LF method provides the true bound. In such situations 
we expect that the method of sections 4.2, 4.4 
can be used to constrain the spectrum of the extremal boundary conditions.
One speculation is the LF bound is the true bound when the bulk CFT
is itself an extremal solution to the bulk modular invariance equations.
It would be interesting to check this by numerical calculations. 

It might be possible to improve the LF method so as to 
produce true lower bounds on $s$ for CFTs such as the $c=2$ 
example or for CFTs with $\Delta_{1}\le (c-1)/12$ where the present  LF method gives no bound at all.
One could try to 
incorporate the constraint that the boundary 
multiplicities $N(h_{j})$ are nonnegative integers. This does not seem possible with the SDP technique,
but one could instead generate 
from the modular transform equation a linear programming problem as was 
done for the three and four-dimensional CFT bootstrap equations \cite{bootstrap_ising}, \cite{extr_func}.
In our case since the modular duality equation \eqref{eq1} contains both positive integer variables $N(h_i)$ and 
real positive variables $b^2(\Delta_k)$ we get a mixed integer linear 
programming (MIP) problem.
Software packages are available for solving such problems. 
A disadvantage of the MIP technique is that unlike the SDP method it 
does not produce  rigorous bounds.
First one has to make assumptions about 
the spectrum of dimensions (in our case about $h_j$) e.g. putting 
them on a grid \cite{bootstrap_ising}, \cite{extr_func}.
Second, one has to have faith in the MIP solver when it says that 
there exists no solution to the MIP problem.  There is no way to 
verify the non-existence.
On a practical level, the extended numerical precision that we have needed 
with the SDP technique is not currently available in  MIP software 
packages.




\section*{Acknowledgements}
CSC thanks Simeon Hellerman and Slava Rychkov for discussions. D.F.\ thanks A.\ Vichi and S.\ El-Showk for helpful discussions of the
linear functional method. The work of CSC was supported by World Premier International Research Center Initiative (WPI Initiative), MEXT, Japan.
The work of AK was supported in part by the STFC grant ST/J000310/1 ``High energy physics at the Tait Institute''. The work of D.F.\ was supported by the Rutgers New High Energy Theory
Center and by U.S.\ Department of Energy Grant No.\ DE-FG02-12ER41813.
\appendix
\renewcommand{\theequation}{\Alph{section}.\arabic{equation}}
\setcounter{equation}{0}
\section{Coefficients $g_{lk}$}
\label{app:glk}
The polynomials
\be
D(z)=\sum_{l=0}^{2n-1} d_{l} z^{l}\,,\qquad
p(x)= \sum_{k=0}^{2n-1}p_{k}x^{k}\,,\qquad
\tilde p(x) = \sum_{k=0}^{2n-1}\tilde p_{k}x^{k}
\ee
are related by equation~\eqref{eq:pptfromD}
\be
p(x)  =x^{-\frac{1}{4}}e^{x/4} D(-4x\partial_{x})\left( x^{\frac{1}{4}}e^{-x/4}\right) \, , \quad 
\tilde p(x) = -x^{-\frac{1}{4}}e^{x/4} D(4x\partial_{x})\left( x^{\frac{1}{4}}e^{-x/4}\right)
\ee
Define polynomials
\be
g_{k}(z) = \sum_{l} g_{lk} z^{l}
\ee
by
\be
x^{k} = x^{-\frac{1}{4}}e^{x/4}  g_{k}(-4x\partial_{x})\left( x^{\frac{1}{4}}e^{-x/4} \right)
\ee
so
\be
D(z) = \sum_{k} p_{k} g_{k}(z)\,,\qquad D(-z) = - \sum_{k} \tilde 
p_{k} g_{k}(z)\,.
\label{eq:Dppt}
\ee
Now calculate
\aeq{
g_{k+1}(-4x\partial_{x})\left( x^{\frac{1}{4}}e^{-x/4} \right)
&=x^{k+1} x^{\frac{1}{4}}e^{-x/4}\\
&=x^{k} x^{\frac{1}{4}}\left(-4x\partial_{x} \right) e^{-x/4}\nonumber\\
&=\left(-4x\partial_{x}+4k+1 \right) x^{k} x^{\frac{1}{4}} e^{-x/4}\nonumber
}
so
\be
g_{0}(z)=1\,,\qquad g_{k+1}(z) = \left(z+4k+1 \right)g_{k}(z)
\ee
so
\be
g_{00}=1\, , \qquad 
g_{0,k+1}=(4k+1)g_{0,k}\, , \qquad 
g_{l,k+1}=g_{l-1,k} + (4k+1)g_{l,k}\, , \enspace 
l=1,2,\dots, k+1\,.
\ee
Equation~\eqref{eq:Dppt} now gives equation~\eqref{eq:constraints}
\be
d_{l} = \sum_{k\ge l} g_{lk} p_{k} = - \sum_{k\ge l} 
(-1)^{l}g_{lk}\tilde p_{k}\,.
\ee

\section{Scaling dimensions in the $c=2$ Gaussian model}
\label{app:c=2}
We need a list of the scaling dimensions  $\Delta_{k}$ of the spin-0 
Virasoro primary fields.
The vertex operators --- the primary fields for the 
$U(1){\times}U(1)$ current algebra ---
have conformal weights
\be
h = p^{2} = p_{1}^{2}+p_{2}^{2}
\,,\qquad
\bar h = \bar p^{2} = \bar p_{1}^{2}+\bar p_{2}^{2}
\ee
\be
p_{\mu}= \frac12(m_{\mu}R+n_{\mu}R^{-1})
\,,\quad
\bar p_{\mu}= \frac12(m_{\mu}R-n_{\mu}R^{-1})
\,,\quad m_{\mu},n_{\mu}\in\Integers
\,,\quad \mu = 1,2\,.
\ee
Let $N_{h,\bar h}$ be the multiplicity of the 
Virasoro representation with weights $h,\bar h$.
The partition function $\mathrm{tr}\left (q^{L_{0}}\bar q^{\bar 
L_{0}}\right )$ (stripped 
of the factor $q^{-c/24}\bar q^{-c/24}$)
\be
\sum_{p,\bar p} \frac{q^{p^{2}} \bar q^{\bar p^{2}}}
{\prod_{n}\left |1-q^{n}\right |^{4} }
=
\frac{|1-q|^{2}+\sum N_{h,0}q^{h}(1-\bar q)
+\sum N_{0,\bar h}
(1- q) q^{\bar h} +\sum_{h,\bar h\ne 0}N_{h,\bar h} q^{h}\bar q^{\bar h}
}
{\prod_{n}\left |1-q^{n}\right |^{2}}\,.
\ee
can be expanded in the 
characters of the $U(1){\times}U(1)$ current algebra (on the left)
or in the characters of the two Virasoro algebras (on the right).
Multiply by the denominator on the right and rearrange to get
\be
\sum_{h,\bar h\ne 0} N_{h,\bar h}q^{h}\bar q^{\bar h}
=
\sum_{p,\bar p} \frac{q^{p^{2}} \bar q^{\bar p^{2}}}
{\prod_{n}\left |1-q^{n}\right |^{2} }
-|1-q|^{2}-\sum N_{h,0}q^{h}(1-\bar q)
-\sum N_{0,\bar h}
(1- q) q^{\bar h}
\label{eq:c=2genfn1}
\ee
Write $P_{s=0}$ for the projection on the spin-0 part of a sum over 
powers of $q$ and $\bar q$ --- the terms with the same power of $q$ 
and $\bar q$, and apply it to both sides of the above identity.
\be
\sum_{h=\bar h\ne 0} N_{h,\bar h}q^{h}\bar q^{\bar h}
=
P_{s=0}
\sum_{p,\bar p} \frac{q^{p^{2}} \bar q^{\bar p^{2}}}
{\left |\prod_{n}(1-q^{n})\right |^{2} }
-1 - q\bar q + N_{1,0}q\bar q + N_{0,1}q\bar q 
\label{eq:c=2genfn}
\ee
Look at the $q$ and $\bar q$ terms in \eqref{eq:c=2genfn1}.
There are no $p,\bar p$ with $p^{2}=1$, $\bar p=0$ or  $p^{2}=0$, 
$\bar p=1$, so
\aeq{
0 &= q + q -N_{1,0}q \\
0 &= \bar q + \bar q -N_{0,1}\bar q\nonumber
}
so $N_{1,0}=N_{0,1}=2$.  Equation~\eqref{eq:c=2genfn} becomes
\be
\sum_{h=\bar h\ne 0} N_{h,\bar h}q^{h}\bar q^{\bar h}
=
P_{s=0}
\sum_{p,\bar p} \frac{q^{p^{2}} \bar q^{\bar p^{2}}}
{\left |\prod_{n}(1-q^{n})\right |^{2} }
-1 +3 q\bar q
\,.
\ee
By inspection, $N_{h,h}=N(\Delta)\ne 0$ exactly for all $\Delta = 2h$ of the form
\be
\Delta = p^{2} + \bar p^{2} + |p^{2} - \bar p^{2}| +2r \,,\qquad 
r=0,1,\ldots
\ee
which is
\be
\Delta = m_{1}^{2} + m_{2}^{2} + \frac14 (n_{1}^{2}+n_{2}^{2})
+|m_{1}n_{1}+m_{2}n_{2}| +2 r \,,\qquad r=0,1,\ldots
\ee
or
\be
4\Delta = 4(m_{1}^{2} + m_{2}^{2} +|m_{1}n_{1}+m_{2}n_{2}|) +  n_{1}^{2}+n_{2}^{2}
 +8 r \,,\qquad r=0,1,\ldots
\ee
Consider the cases
\be
(m_{1},m_{2},n_{1},n_{2}) = 
(0,0,0,0),\,(0,0,1,0),\,(0,0,1,1),\,(1,0,0,0),\,(1,0,0,1)
\ee
to get
\be
N(\Delta) \ne 0 \text{ for } 4\Delta \equiv 0,1,2,4,5\,(\mathrm{mod} \,8)
\ee
Finally, we show that $N(\Delta)=0$ for $4\Delta \equiv 3,6,7\,(\mathrm{mod}\, 8)$ which is to say
for (1) $4\Delta \equiv -1 \,(\mathrm{mod}\, 4)$, and
(2) $4\Delta \equiv 6\,(\mathrm{mod} \,8)$.

For (1), note that $4\Delta \equiv n_{1}^{2}+n_{2}^{2}\,(\mathrm{mod}\, 4)$.  If $n_{1}$ and $n_{2}$ are 
both even, then $4\Delta \equiv 0\,(\mathrm{mod}\, 4)$.  If both are odd, $4\Delta \equiv 2 
\,(\mathrm{mod}\, 4)$.  If one is even and the other is odd, $4\Delta \equiv 1 \,(\mathrm{mod}\, 4)$.
So $4\Delta \not\equiv -1 \,(\mathrm{mod}\, 4)$. 

For (2), suppose that $4\Delta \equiv 6\,(\mathrm{mod} \,8)$ then 
$4\Delta \equiv 2\,(\mathrm{mod}\, 4)$ so $n_{1}$ and 
$n_{2}$ must both be odd, say $n_{1}=2k_{1}+1$, $n_{2}=2k_{2}+1$.  
Then
\be
4\Delta = 4(m_{1}^{2} + m_{2}^{2} 
+|2m_{1}k_{1}+m_{1}+2m_{2}k_{2}+m_{2}|) +  4 k_{1}(k_{1}+1) + 4 k_{2}(k_{2}+1)  +2
 +8 r
\ee
so $m_{1}^{2} + m_{2}^{2}  +|2m_{1}k_{1}+m_{1}+2m_{2}k_{2}+m_{2}|$ 
must be odd, so $m_{1}^{2} + m_{2}^{2}+m_{1}+m_{2} $ must be odd, 
which is impossible.
So $4\Delta \not\equiv 6 \,(\mathrm{mod}\, 8)$. 

Therefore
\be
\{\Delta_{k} \} = \left \{ \frac{m}4: m>0, m\equiv 
0,1,2,4,5\,(\mathrm{mod} 8)\right \}\,.
\ee

\end{document}